\documentclass[11pt]{article}
\usepackage{graphicx}

\begin{document}

\def\xslash#1{{\rlap{$#1$}/}}
\def \p {\partial}
\def \dd {\psi_{u\bar dg}}
\def \ddp {\psi_{u\bar dgg}}
\def \pq {\psi_{u\bar d\bar uu}}
\def \jpsi {J/\psi}
\def \psip {\psi^\prime}
\def \to {\rightarrow}
\def\bfsig{\mbox{\boldmath$\sigma$}}
\def\DT{\mbox{\boldmath$\Delta_T $}}
\def\xit{\mbox{\boldmath$\xi_\perp $}}
\def \jpsi {J/\psi}
\def\bfej{\mbox{\boldmath$\varepsilon$}}
\def \t {\tilde}
\def\epn {\varepsilon}
\def \up {\uparrow}
\def \dn {\downarrow}
\def \da {\dagger}
\def \pn3 {\phi_{u\bar d g}}

\def \p4n {\phi_{u\bar d gg}}

\def \bx {\bar x}
\def \by {\bar y}

\begin{center}
{\Large\bf Partonic State and Single Transverse Spin Asymmetry in Drell-Yan Process} \vskip 10mm
J.P. Ma and H.Z. Sang    \\
{\small {\it  Institute of Theoretical Physics, Academia Sinica, P.O. Box 2735,
Beijing 100080, China }} \\
\end{center}
\vskip 10mm
\begin{abstract}
Single transverse-spin asymmetries have been studied intensively both in experiment and theory.
Theoretically, two factorization approaches have been proposed. One is by using transverse-momentum-dependent
factorization and the asymmetry comes from the so called Sivers function. Another is by using collinear factorization
where the nonperturbative effect is parameterized by a twist-3 hadronic matrix element. However,
the factorized formulas for the asymmetries in the two approaches are derived at hadron level
formally by diagram expansion, where one works with various parton density matrices of hadrons.
If the two factorizations hold, they should also hold at parton level. We examine this for Drell-Yan
processes by replacing hadrons with partons. By calculating the asymmetry, Sivers function and the twist-3
matrix element at nontrivial leading order of $\alpha_s$, we find that we can reproduce
the result of the transverse-momentum-dependent factorization. But we can only verify
the result of the collinear factorization partly. 
Two formally derived relations between Sivers function
and the twist-3 matrix element are also examined with negative results.
\end{abstract}
\vskip 1cm

\par
\noindent
{\bf 1. Introduction}
\par
Single transverse-spin asymmetries have been observed in many experiments\cite{E704,hermes,SMC,COM,STPH}.
To generate a single transverse-spin asymmetry(SSA) it requires nonzero absorptive part of scattering amplitude
and helicity-flip interactions. This indicates that the asymmetries are $T$-odd effects.
The study of SSA of hadron scattering provides a new tool to explore
hadron structure and nonperturbative properties of QCD.
In some cases like production of a heavy quark with transverse polarization,
SSA can be studied by using perturbative QCD directly\cite{TPL}.
However, in the cases studied in experiment, in which an initial hadron is transversely polarized and
is involved in the scattering, it is not possible to use perturbative QCD directly.
To study SSA of hadronic processes
two factorization approaches have been proposed. One is by using
transverse-momentum-dependent(TMD) factorization, where one takes transverse momenta of partons in
hadrons into account. Another is the collinear factorization. An up-to-date review about
studies of SSA can be found in \cite{Review}. In this work we will focus on
the two approaches in Drell-Yan processes.
\par
In the approach of TMD factorization, the origin of SSA arises from a correlation between
the transverse spin of the initial hadron and the transverse momentum of partons in the hadron.
This correlation is parameterized by Sivers function\cite{Sivers,JC}. In this approach
the effect of helicity-flip interactions and
the $T$-odd effect are contained in the Sivers function.
The helicity-flip
of a initial hadron can happen because of orbital angular momenta
of partons. This can be seen clearly in terms of light-cone wave functions\cite{JMYW}.
Hence SSA in this approach is sensitive to orbital angular momenta
of partons.
The $T$-odd effect in Drell-Yan
processes comes from the initial state interaction, in contrast to semi-inclusive DIS where
$T$-odd effects come from final state interactions.
We note here
that so far TMD factorization has been examined carefully only for physical quantities
which do not contain $T$-odd effects\cite{CS,CSS,JMY,JMYG,CAM}. TMD parton distributions
entering the factorization for these physical quantities can be defined with QCD operators
consistently. Intensive efforts in theory has been spent
to study how to consistently define or interpret Sivers function as a parton distribution
which is gauge invariant and contains initial- or final state interactions\cite{JC,SJ1,TMDJi,Mulders97,Boer03}.
Through these studies the role of gauge links used to define Sivers function becomes clear
and it shows that the Sivers function in Drell-Yan processes is related to that in semi-inclusive DIS.
The approach of TMD factorization has a simple parton-model interpretation. Because
of this,
SSA has been studied extensively in terms of Sivers functions
\cite{Anselmino,Mulders,DeSanctis,Efremov,BQMa,Liang}.
These functions have been also studied with models\cite{MCollins,MSivers1,MSivers2,Fyuan}.
The approach of TMD factorization has the limitation that it is only applicable
in certain kinematic regions, e.g., in a Drell-Yan process the region is where
the transverse momentum $q_\perp$ of the lepton pair is much smaller
than its invariant mass $Q$.
\par
In the approach of collinear factorization SSA  is factorized with twist-3 matrix elements\cite{QiuSt, EFTE,KaKo},
or called ETQS matrix elements. In this approach the twist-3 matrix elements, or the corresponding parton
distribution functions defined with twist-3 operators of QCD, contain only the effect of helicity-flip interactions
which is taken as nonperturbative effect. The nonzero absorptive part or $T$-odd effect
is not contained in the twist-3 matrix elements. It is generated by poles of parton propagators in hard scattering.
The twist-3 matrix elements characterize the correlation between quarks and gluons inside the transversely
polarized hadron.
Therefore, in this approach SSA is sensitive to the correlation.
From this point of view the approach of
collinear factorization seems different than the approach of TMD factorization. However,
recent progress shows that the two approaches can be unified in the kinematic region
of $q_\perp << Q$\cite{JQVY}. We note here that the approach of collinear factorization
is applicable for the whole kinematical region if $Q^2$ is enough large.
The fact that the two approaches
in the region of $q_\perp << Q$ indicates that there exists a relation between
Sivers function and twist-3 matrix elements. Such a relation has been found in
\cite{JQVY}. There also exists another relation between
Sivers function and twist-3 matrix elements\cite{Boer03,MW1}.
Applications of the collinear factorization for SSA can be found in \cite{tw31,tw32,tw33}.
\par
It should be noted that in the two approaches the factorization is derived or proposed rather formally in
the sense that one works at hadron level by using the diagram expansion.
In the expansion one usually divides a given diagram with hadrons into various parts.
Among these various parts, one consists only of partons. In other parts hadrons are involved.
These parts represent nonperturbative effects related to the hadrons and they are parameterized
by various parton density
matrices of hadrons. It should be also noted that QCD factorizations,  if they are proven, are
general properties of QCD Green functions. It means that the two factorization approaches,
if they hold, they should also hold by replacing hadrons with partons. It is the purpose
of the study presented here to show how SSA in Drell-Yan processes can be factorized in two
approaches by replacing hadrons with partons and to examine the two relations between
Sivers function and twist-3 matrix elements. Our study is performed at leading order
of $\alpha_s$. In order to generate SSA in Drell-Yan processes and nonzero
$q_\perp$, there must be exchange of two gluons at the leading order. It results in that
SSA at parton level is already at order of $\alpha_s^2$ in comparison with the leading order
of the unpolarized cross-section which is at $\alpha^0_s$.
Hence it is nontrivial to show the factorizations at leading order.
In this work we will take Drell-Yan process as an example. We replace the two hadrons in the initial state
with a quark $q$ and an antiquark $\bar q$. In order to have helicity-flip we keep
the quark mass as nonzero and every quantity is calculated at leading power of $m$.
The perturbative coefficients in the factorization formulas do not depend on the quark mass $m$.
It turns out that the proposed TMD factorization of SSA holds at the parton level, while
only a part of results of the proposed collinear factorization for SSA can be
verified with our partonic results. We also find that the two relations
between Sivers functions and twist-3 matrix elements do not hold in general.
The two relations need to be modified.
\par
Our paper is organized as the following: In Sect.2 we give our notations for
Drell-Yan process and the formulas of two factorization approaches of SSA.
In Sect.3 we present our result of Sivers function with a parton state.
SSA of Drell-Yan process with the parton state is calculated in Sect.4.
Sect.5 contains the result of twist-3 matrix element with the parton state,
where we show that only a part of our result matches the formula of the collinear
factorization of SSA.
We summarize our study in Sect.6.

\par\vskip20pt
\noindent
{\bf 2. Notations and Factorization Formulas}
\par
We consider the Drell-Yan process:
\begin{equation}
  h_A ( P_A, s) + h_B(P_B) \to \gamma^* (q) +X \to  \ell^-  + \ell ^+  + X,
\end{equation}
where $h_A$ is a spin-1/2 hadron with the spin-vector $s$.
We will use the  light-cone coordinate system, in which a
vector $a^\mu$ is expressed as $a^\mu = (a^+, a^-, \vec a_\perp) =
((a^0+a^3)/\sqrt{2}, (a^0-a^3)/\sqrt{2}, a^1, a^2)$ and $a_\perp^2
=(a^1)^2+(a^2)^2$.
We also introduce two light-cone vectors: $n^\mu =(0,1,0,0)$ and $l^\mu = (1,0,0,0)$.
We take a light-cone coordinate system in which the momenta and the spin are :
\begin{equation}
P_{A,B}^\mu = (P_{A,B}^+, P_{A, B}^-, 0,0),  \ \ \ \  s^\mu =(0,0, \vec s_\perp).
\end{equation}
$h_A$ moves in the $z$-direction, i.e., $P_A^+$ is the large component. The spin of $h_B$
is not observed. The invariant mass of the observed lepton pair is $Q^2 =q^2$.
The relevant hadronic tensor is defined as:
\begin{equation}
W^{\mu\nu}  = \sum_X \int \frac{d^4 x}{(2\pi)^4} e^{iq \cdot x} \langle h_A (P_A, s_\perp), h_B(P_B)  \vert
    \bar q(0) \gamma^\nu q(0) \vert X\rangle \langle X \vert \bar q(x) \gamma^\mu q(x) \vert
     h_B(P_B),h_A (P_A, s_\perp)  \rangle,
\end{equation}
and the differential cross-section is determined by the hadronic tensor as:
\begin{equation}
\frac{ d\sigma }{ dQ^2 d^2 q_\perp d q^+ d q^- } = \frac{4\pi \alpha_{em}^2 Q_q^2}{3 S Q^2}
    \delta (q^2 -Q^2)
    \left ( \frac {q_\mu q_\nu} {q^2} - g_{\mu\nu} \right ) W^{\mu\nu} .
\end{equation}
\par
We are interested in the kinematical region where $q_\perp^2 << Q^2$. The hadronic tensor
at leading twist accuracy has the structure:
\begin{eqnarray}
W^{\mu\nu} &=& - g_\perp^{\mu\nu} W_U^{(1)} + \left ( g_\perp^{\mu\nu}
-2 \frac{q_\perp^\mu q_\perp^\nu} {q_\perp^2}  \right )
   W_U^{(2)}
\nonumber\\
&&  - g_\perp^{\mu\nu} \epsilon_\perp^{\alpha \beta} s_{\perp\alpha} q_{\perp\beta} W_T^{(1)}
 + \left ( s_{\perp\alpha} \epsilon_\perp^{\alpha\mu} q_\perp^\nu
          +s_{\perp\alpha} \epsilon_\perp^{\alpha\nu} q_\perp^\mu -g_\perp^{\mu\nu}
             \epsilon_\perp^{\alpha \beta} s_{\perp\alpha}  q_{\perp\beta} \right ) W_T^{(2)}
\nonumber\\
  &&  +  q_{\perp\alpha} \left ( \epsilon_\perp^{\alpha\mu}  q_{\perp}^\nu
 + \epsilon_\perp^{\alpha\nu} q_{\perp}^\mu \right ) {\vec  q}_\perp \cdot \vec s_\perp W_T^{(3)}
 +\cdots
\end{eqnarray}
with the notation:
\begin{equation}
  g_\perp^{\mu\nu} = g^{\mu\nu} - n^\mu l^\nu - n^\nu l^\mu,
  \ \ \ \ \ \
  \epsilon_\perp^{\mu\nu} =\epsilon^{\alpha\beta\mu\nu}l_\alpha n_\beta.  \ \ \ \
\end{equation}
In the above, we only give the structures symmetric in $\mu\nu$. $W_T^{(i)}(i=1,2,3)$ represent
$T$-odd effect related to the spin. $W_U^{(1,2)}$ are responsible for unpolarized cross-sections.
$W_T^{(1)}$ contributes to SSA in the region $q^2_\perp << Q^2$
\begin{equation}
\frac{ d\sigma (\vec s_\perp )}{ dQ^2 d^2 q_\perp d q^+ d q^- }
 -\frac{ d\sigma (- \vec s_\perp ) }{ dQ^2 d^2 q_\perp d q^+ d q^- }
  =\frac{16\pi \alpha_{em}^2 Q_q^2}{3 S Q^2}\delta (q^2 -Q^2) \epsilon_\perp^{\alpha \beta} s_{\perp\alpha} q_{\perp\beta} W_T^{(1)}
   \left ( 1 + {\mathcal O} ( q^2_\perp / Q^2) \right ).
\end{equation}
We will focus on $W_T^{(1)}$ to see if it can be factorized.
\par
In the kinematical region $Q^2>> q^2_\perp \sim \Lambda_{QCD}^2$ the TMD factorization can be performed
for $W_T^{(1)}$ based on the diagram expansion. The result at tree-level can be written
as a convolution with Sivers function and TMD parton distribution\cite{JMY,JQVY}:
\begin{eqnarray}
W^{(1)}_T (z_1,z_2,q_\perp)
      = \frac{1}{N_c}
  \int d^2 k_{1\perp}  d^2 k_{2\perp} \frac{\vec q_\perp \cdot \vec k_{1\perp}}{q^2_\perp}
  q_\perp (z_1, k_{1\perp}) \bar q(z_2, k_{2\perp})
    \delta^2 (\vec k_{1\perp} +\vec k_{2\perp} -\vec q_\perp ) H ,
\label{TMDFAC}
\end{eqnarray}
where the
variables $z_{1,2}$ are defined as: $q^+ = z_1 P_A^+$ and $q^- = z_2 P_B^-$.
$H$  is a perturbative coefficient, i.e., $H = 1 + {\mathcal O}(\alpha_s)$.
Beyond tree-level one has to implement a soft factor representing effects of soft-gluon radiation.
In the above $q_\perp$ is Sivers function. To define it with QCD operators we introduce a gauge link
along the direction $u$ with $u^\mu =(u^+,u^-,0,0)$:
\begin{equation}
L_u (-\infty, z) = \left [ P \exp \left ( -i g_s \int_{-\infty}^0  d\lambda
     u\cdot G (\lambda u + z) \right ) \right ] ^\dagger .
\end{equation}
The Sivers function relevant for Drell-Yan process is defined in the limit $u^+ << u^-$
\cite{JC,JMY,Mulders97}:
\begin{eqnarray}
q_\perp (x,k_\perp)
\varepsilon_\perp^{\mu\nu} s_{\perp\mu}  k_{\perp\nu}
  & =& \frac{1}{4}  \int \frac{dz^- d^2 z_\perp}{(2\pi)^3}
                e ^{-i k \cdot z }
               \{
                \langle P_A, \vec  s_\perp \vert
                \bar\psi (z ) L_u ^\dagger (-\infty, z) \gamma^+
                L_u (-\infty, 0) \psi(0) \vert P_A,\vec s_\perp \rangle
 \nonumber\\
         &&          - (\vec s_\perp \to -\vec s_\perp ) \} ,
\end{eqnarray}
with $z^\mu =(0,z^-,\vec z_\perp)$. $x$ is defined as $k^+ = xP^+_A $.
Beside the renormailzation scale $\mu$, Sivers function  also depends  on the parameter:
\begin{equation}
\zeta^2 = \frac {2 u^-}{u^+} \left ( P^+ \right )^2.
\end{equation}
The limit $u^+ << u^-$ is to be understood that we discard all contributions in Eq.(9) which
are zero with $\zeta^2 \to \infty$. The definitions of other TMD parton distributions of
a unpolarized hadron can be found in \cite{JMY,JMYG}.
In the TMD factorization of SSA one takes transverse momenta of incoming partons
into account. The $T$-odd effect and spin-flip effect are parameterized
by Sivers function. It should be noted that the TMD factorization can be extended
to the region $Q^2 >> q^2_\perp >> \Lambda_{QCD}^2$.
\par
A collinear factorization can also be performed for SSA or $W_T^{(1)}$, where the $T$-odd effect
comes from poles of partons in the hard scattering and the spin-flip effect is parameterized
with the twist-3 matrix element which is defined as\cite{QiuSt,EFTE}:
\begin{eqnarray}
T_F (x_1,x_2)
s_{\perp}^\nu
   & =&    \frac{g_s}{2}\int \frac{dy_1 dy_2}{4\pi}
   e^{ -iy_2 (x_2-x_1) P^+ -i y_1 x_1 P^+ } \epsilon_\perp^{\mu\nu}
\nonumber\\
    && \cdot \left  \{ \langle P_A, \vec s_\perp \vert
           \bar\psi (y_1n ) \gamma^+ G^{+}_{\ \ \mu}(y_2n) \psi(0) \vert P_A,\vec s_\perp \rangle
  - ( \vec s_\perp \to - \vec s_\perp ) \right \}.
\label{tw3}
\end{eqnarray}
In the above we have suppressed gauge links between operators. These gauge links are defined
with the vector $n$ and make the above definition gauge invariant. One can also view the definition
as given in the gauge $n\cdot G=0$. With the twist-3 matrix element
SSA or $W_T^{(1)}$ takes the following factorized form\cite{JQVY}:
\begin{equation}
 W_T^{(1)} \sim \bar q \otimes H_c \otimes T_F,
\label{CF}
\end{equation}
where $\bar q$ is the standard parton distribution, $H_c$ is a coefficient function calculated
perturbatively. Its leading order is at $\alpha_s$. Details about the above result can be found
in \cite{JQVY}.
It should be noted that the above collinear factorization is derived
for the kinematical region with $Q^2 >> \Lambda^2_{QCD}$ and $q^2_\perp >> \Lambda^2_{QCD}$.
It should also be valid for the region with
$\Lambda^2_{QCD}<<q^2_\perp<< Q^2$. In this region the two factorization approaches apply.
It has been shown both factorizations give the same results\cite{JQVY} in that region.
Hence a relation between Sivers function
$q_\perp$ and the twist-3 matrix element $T_F$ can be found.
\par
As discussed in the introduction the derivation of these factorization formulas is based on
the diagram expansion, in which one works with hadronic states by introducing
various parton density matrices of hadrons. If these factorization formulas
hold, it should also hold if we replace hadrons with partons.
It is the task of the subsequent sections to check
these factorizations and different relations between Sivers function and the twist-3 matrix
element by replacing hadrons with partons.

\par\vskip20pt
\par
\begin{figure}[hbt]
\begin{center}
\includegraphics[width=12cm]{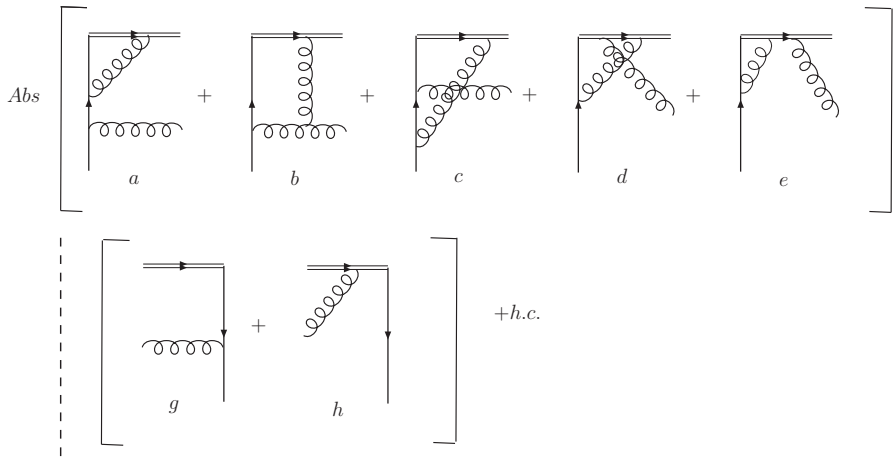}
\end{center}
\caption{Diagrams for the contributions to Sivers function. The double lines represent
the gauge link. }
\label{Feynman-dg1}
\end{figure}
\par
\noindent
{\bf 3. Sivers Function with a Quark-State}
\par
We replace in the definition of Sivers function the hadron $h_A$ with a quark $q$.
The quark has the momentum $p^\mu =(p^+, p^-,0,0)$ with $p^2 =m^2$. The finite quark mass
$m$ will introduce effects of spin-flip.
In order to have $T$-odd effect, exchanges of virtual gluons should present in the amplitude.
Also, at least one gluon must be in the intermediate state to generate nonzero $k_\perp$.
With these considerations one can find the possible contributions to Sivers function
at leading order of $\alpha_s$. These contributions are given by diagrams
in Fig.1., where the contributions are represented as the interference of amplitudes.
The amplitudes of Fig.1a to Fig.1e can have absorptive parts. The interference of these absorptive parts
with the amplitudes represented with Fig.1g and Fig.1h will give nonzero contributions
to Sivers function.
\par
\begin{figure}[hbt]
\begin{center}
\includegraphics[width=12cm]{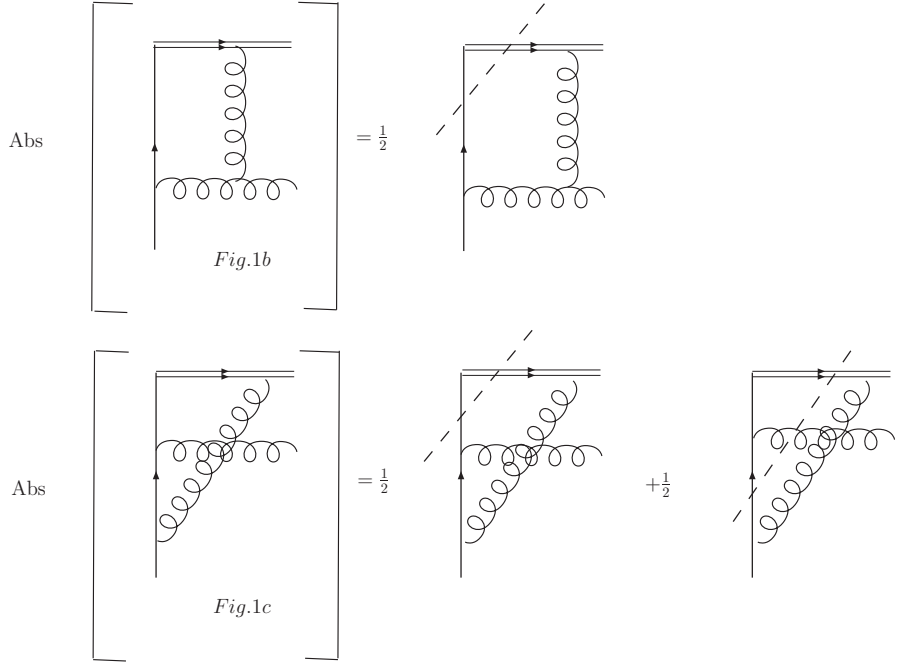}
\end{center}
\caption{The absorptive part of Fig.1b and Fig.1c. The broken line is the cut. }
\label{Feynman-dg1}
\end{figure}
\par
At first look there are many diagrams contributing.
However, the Sivers function $q_\perp (x,k_\perp)$ is defined in the limit $u^-<< u^+$.
In this limit one can easily show that only the interference of Fig.1b and Fig.1c with Fig.1g
are nonzero, other interferences are zero.
The absorptive parts of amplitudes can be obtained with the standard Cutkosky cutting rule
and they can be represented with cut diagrams. In Fig.2 we give the diagrams for
the absorptive part of Fig.1b and Fig.1c. In determining physical cuts one should note
that the energy flow of each particle crossing the cut should be in the same direction.
In our case, the gauge link represents an incoming particle with an infinity-large $-$-component
of its momentum. Therefore, the energy flow of the gauge links in Fig.2
are from the right side  to the left side. Keeping this in mind we find
only one cut diagram for Fig.1b with a physically allowed cut.
For Fig.1c there are two possible cuts. A cut cutting a real particle has no effect.
According to the Cutkosky cutting rule one should replace all $i$'s of propagators and vertices
with $-i$ in the left-upper part of the cut diagrams in Fig.2. In this part of the cut diagrams
the joining point of the gauge link and the quark line should be taken as a vertex, reflecting
the fact that the particle represented by the gauge link annihilates the particle represented
by the quark line. Hence, this joining point or vertex contributes an extra minus sign when evaluating
the absorptive part of the cut diagrams. This extra minus sign can also be verified by a direct calculation
of the interference of the complex conjugated Fig.1b or Fig.1c with the complex conjugated Fig.1g.
\par
The interference of Fig.1b with Fig.1g can be written as:
\begin{eqnarray}
q_\perp (x,k_\perp)
\varepsilon_\perp^{\mu\nu} s_{\perp\mu}  k_{\perp\nu}\vert_{bg}   &=&  -\frac{1}{4N_c} g_s^4 f^{abc} \frac{1}{2}
 \int \frac {d^4 k_g}{(2\pi)^4} \frac {d^4 q}{(2\pi)^4}
   (2\pi) \delta (k^2_g) \delta ( k^+ -p^+ + k_g^+) \delta^2 ( \vec k_\perp + \vec k_{g\perp} )
\nonumber\\
  && \cdot \frac{-i2\pi \delta (u\cdot q)}{(p-k_g)^2 -m^2 -i\varepsilon} \cdot
  \frac{-i2 \pi \delta( (p-k_g-q)^2 -m^2)}{(q+k_g)^2 + i\varepsilon} \cdot \frac{1}{q^2 +i\varepsilon}
\nonumber\\
    && \cdot \bar u(p,s_\perp) \gamma_\mu (\gamma\cdot (p-k_g) +m) \gamma^+
    (\gamma\cdot (p-k_g-q) +m) \gamma_\rho T^aT^b T^c u(p,s_\perp)
\nonumber\\
    && \cdot \left [ (-k_g +q)^\rho u^\mu +(-2q-k_q)^\mu u^\rho + (2k_g+q)\cdot u g^{\mu\rho} \right ]
      -\left ( s_\perp \to -s_\perp \right ),
\end{eqnarray}
where $k_g$ is the momentum of the gluon in the intermediate state and $q$ the momentum of the gluon
emitted from the gauge link.
It is straightforward to calculate the contribution under the limit $u^+ \to 0$.
We have:
\begin{equation}
 q_\perp(x,k_\perp)\vert_{bg}  =  \frac{m \alpha_s^2}{16 \pi^2}  (N_c^2-1)  \frac{ x(1-x)}{ k^2_\perp + (1-x)^2 m^2}
\nonumber\\
  \frac{1}{ k^2_\perp } \ln
    \left ( \frac  {(1-x)^2 m^2}{k^2_\perp +(1-x)^2 m^2 }\right ).
\end{equation}
\par
The absorptive part of Fig.1c receives contributions from two cut diagrams in Fig.2. It is easy to show
that the contributions from the two cut diagrams cancel each other. Therefore the only nonzero contribution
is from Fig.1b.
The final result of Sivers function is:
\begin{eqnarray}
 q_\perp (x, k_\perp )  =  \frac{m \alpha_s^2}{8 \pi^2}  (N_c^2-1)  \frac{ x(1-x)}{ k^2_\perp + (1-x)^2 m^2}
 \frac{1}{ k^2_\perp } \ln
    \left ( \frac{(1-x)^2 m^2}{k^2_\perp +(1-x)^2 m^2 }  \right ).
\label{Sivers}
\end{eqnarray}
\par
We note that the same diagrams in Fig.1
will also contribute to Sivers function in DIS, where the gauge link
is pointing to the future. This gauge link then represents
an outgoing particle with an infinity-large $-$-component
of its momentum, hence the energy flow along the gauge link is reversed.
This will lead to cut diagrams other than those given in Fig.2. It turns out
that Sivers function for DIS calculated with Fig.1 is the same as the above,
except a sign difference as expected. We also point out that
there can be a difference in calculations of Sivers function or SSA
between different ways to generate absorptive parts.
In the collinear expansion imaginary parts of amplitudes are generated
by poles of parton propagators. In general the generated imaginary parts
can not be the absorptive parts, because these poles do not necessarily correspond
to physical cuts,  although the same results may be obtained. The difference
and SSA in DIS will be studied in a separate publication.

\par\vskip20pt
\noindent {\bf 4. SSA of Drell-Yan Processes}
\par
In this section we will calculate the part of the hadronic tensor relevant to SSA.
We replace $h_A$ with a quark and $h_B$ with an antiquark. We consider SSA in
the process:
\begin{equation}
    q(p_1,s) + \bar q(p_2) \to \gamma^* (q) + X \to  \ell^-  + \ell ^+  + X ,
\end{equation}
where the quark $q$ is polarized with the spin vector $s$. By the requirement
that there is a $T$-odd effect and nonzero $q_\perp$, we find that at leading order
of $\alpha_s$
the possible contributions to SSA are given by diagrams in Fig.3. These diagrams
are of the hadronic tensor and the black dot indicates the insertion of the
electric current. We denote
the momentum of the gluon in the intermediate state as $k_g$.
It will be very tedious to obtain full results from these diagrams. However, what we need
is the leading contribution in the limit $q^2_\perp << Q^2$. It will be useful
by doing the expansion in $q^2_\perp /Q^2$ first and then to perform the loop integral.
A convenient way for the expansion is to analysis different regions of
the loop momentum.
\par
\begin{figure}[hbt]
\begin{center}
\includegraphics[width=12cm]{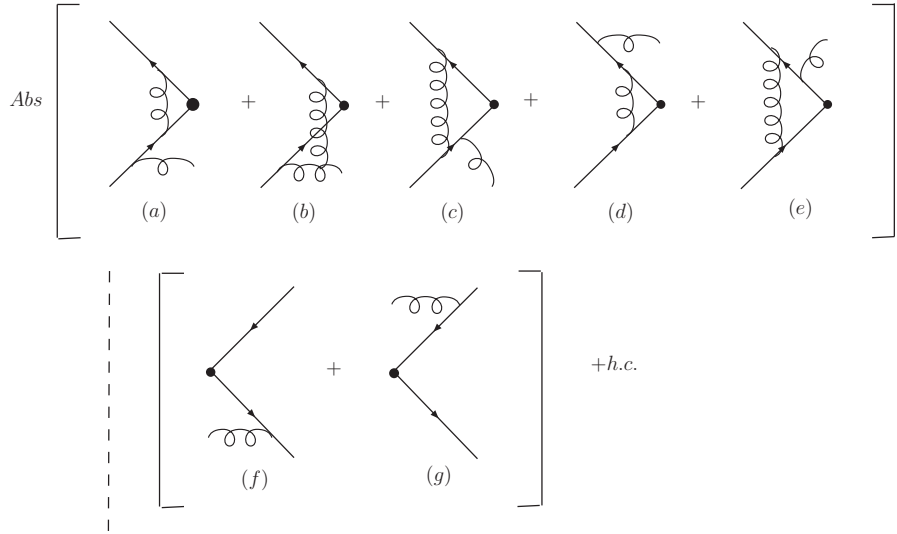}
\end{center}
\caption{Diagrams for the contributions to the hadronic tensor $W_T^{(1)}$. }
\label{Feynman-dg5}
\end{figure}
\par
For doing the expansion we note that each contribution from Fig.3 can be written
in a generic form
\begin{equation}
\int \frac{d^4 k}{(2\pi)^4} \frac{ d^4 k_g} {(2\pi)^4} (2\pi) \delta (k_g^2)
               (2\pi)^4 \delta^4(p_1 +p_2 -q -k_g )
     \frac{1}{D_1 D_2 D_3 D_4 D_5 } \cdot {\rm Tr} \left [ \cdots \right ],
\end{equation}
where $k$ is the momentum of the virtual gluon. In each contribution there are
five propagators, their denominators are denoted as $D_i(i=1,2,\cdots, 5)$.
The nominator represented in the above as ${\rm Tr} \left [ \cdots \right ]$
is a trace of product of $\gamma$-matrices.
We scale the momentum $\vec k_{g\perp} =-\vec q_\perp$ as at order of $\lambda$.
and expand each contribution in $\lambda$.
It is clearly that
the leading order contribution in $\lambda$ comes when the denominators of all
propagators are at order of $\lambda^2$.
The power-counting is not affected by taking a cut to cut propagators.
The requirement that all denominators in Fig.3a to Fig.3e are at order
of $\lambda^2$ gives the following scaling of loop momenta:
\begin{equation}
  k^\mu = (\lambda^2, \lambda^2, \lambda, \lambda), \ \ \ \ \ k_g^\mu =( k_g^+, k_g^-, \lambda, \lambda),
\end{equation}
where one of the component $k_g^+$ or $k_g^-$ is at order of $\lambda^2$. It depends on diagrams. However,
some contributions can not have all denominators at order of $\lambda^2$, e.g.,
the interference terms of Fig.3a or Fig.3c with Fig.3g, the interference terms of
Fig.3d or Fig.3e with Fig.3f. It is easy to find a rule to determine which contribution
is dominant. If $k_g^-$ in one of Fig.3a to Fig.3e is at order of $\lambda^2$, then its interference
with Fig.3g is not at leading order of $\lambda$. If $k_g^+$ in one of Fig.3a to Fig.3e is at order of $\lambda^2$,
then its interference
with Fig.3f is not at leading order of $\lambda$. We note that the leading order of the nominators
starts at order of ${\mathcal O}(\lambda)$ or higher.
With this rule we can only have the leading contributions from the following interference terms:
Fig.3a or Fig.3c with Fig.3f, Fig.3d or Fig.3e with Fig.3g, Fig.3b with Fig.3f if $k_g^-$ is at order
of $\lambda^2$, and Fig.3b with Fig.3g if $k_g^+$ is at order of $\lambda^2$. In these
contributions all denominators of propagators are at order of $\lambda^2$. Evaluating the nominator
of these contributions we find that only the nominator of Fig.3b interfered with Fig.3f and
that of Fig.3a interfered with Fig.3f are at order of $\lambda$, other nominators are at order
of $\lambda^3$. Therefore the leading order contributions come only from the interference
term of Fig.3b or Fig.3c with Fig.3f. We note that the leading contribution comes from
the Glauber region of the momentum of the virtual gluon. Hence its propagator can be approximated
as:
\begin{equation}
   \frac{i}{k^2+ i\varepsilon} \approx \frac{i}{-k^2_\perp + i\varepsilon}.
\end{equation}
From the above analysis, one can see that the diagram Fig.3b and Fig.3c, which can generate nonzero SSA,
are in correspondence to the diagram Fig.1b and Fig.1c, which contribute to Sivers function, respectively.
\par
\par
\begin{figure}[hbt]
\begin{center}
\includegraphics[width=10cm]{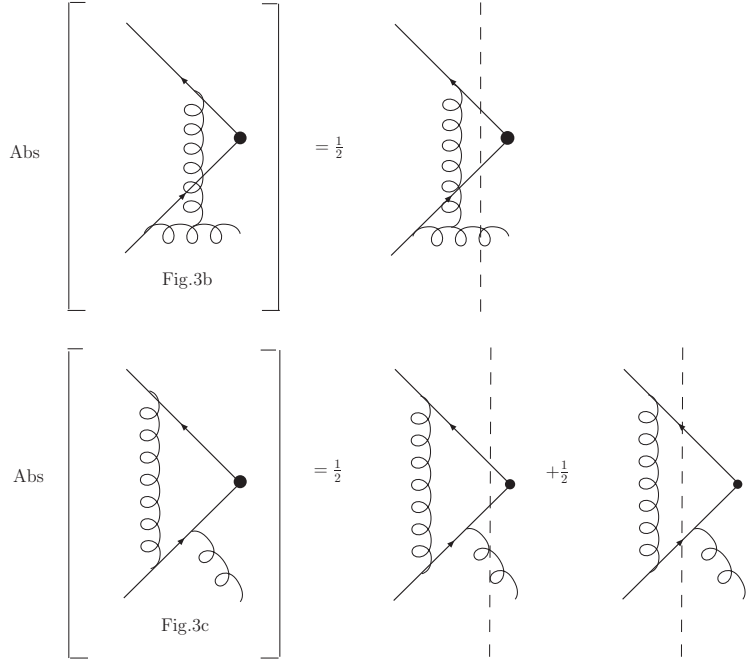}
\end{center}
\caption{The cut diagrams of Fig.3b and Fig.3c for their absorptive parts.}
\label{Feynman-dg5}
\end{figure}
\par
Again, the absorptive parts of Fig.3b and Fig.3c can be found with the Cutkosky cutting rule.
They can be represented by the cut diagrams in Fig.4. For Fig.3b there is only one physically allowed cut,
for Fig.3c there are two.
Taking the interference of Fig.3b with Fig.3f as an example, we have the contribution to the relevant hadronic
tensor as:
\begin{eqnarray}
W^{\mu\nu} \vert_{bf} &=& \frac{g_s^4}{2 N_c^2} \left ( f^{abc} {\rm Tr } T^a T^b T^c \right ) \frac{1}{2 }
              \int \frac{d^4 k}{(2\pi)^4} \frac{ d^4 k_g} {(2\pi)^4} (2\pi) \delta (k_g^2)
               \delta^4(p_1 +p_2 -q -k_g )
\nonumber\\
      &&\cdot  \frac{1}{k^2+ i\varepsilon}\cdot \frac{-i2\pi\delta((p_1-k-k_g)^2 -m^2)}{(k+k_g)^2+ i\varepsilon}
     \cdot \frac{-i2\pi \delta ((p_2+k)^2 -m^2)}{(p_1-k_g)^2 -m^2+ i\varepsilon}
\nonumber\\
     && \cdot \left \{ {\rm Tr} \left [ \frac{1}{2} \gamma_5 \gamma\cdot s_\perp
      (\gamma\cdot p_1 +m) \gamma_\alpha (\gamma\cdot (p_1-k_g) +m)
          \gamma^\nu (\gamma\cdot p_2 -m) \gamma_\beta (-\gamma\cdot (p_2+k) +m) \gamma^\mu
\right.\right.
\nonumber\\
    &&\left.\left.   (\gamma\cdot (p_1-k-k_g)+m)\gamma_\rho \right ]
      \cdot \left [ (-k_g+k)^\rho g^{\alpha\beta} + (-2 k- k_g)^\alpha g^{\beta\rho}
          +(2k_g +k)^\beta g^{\rho\alpha} \right ]\right \}.
\end{eqnarray}
It should noted that one should take complex conjugation of the right part of a cut diagram in evaluating
the absorptive part. With this in mind the black-dotted vertex in the cut diagram of Fig.3b
contributes an extra minus sign. Now it is straightforward to perform the expansion in $\lambda$ and
to pick up the leading contribution. We have then:
\begin{eqnarray}
W^{\mu\nu} \vert_{bf} &=&-g_\perp^{\mu\nu} m \frac{4 \alpha_s}{\pi}\frac{N_c^2-1}{ N_c} 4\pi \delta (k_g^2)
            \frac{(p_2^-)^2 k_g^+ (k_g^+ -p_1^+) }{(p_1-k_g)^2 -m^2}
            \epsilon_\perp^{\alpha \beta}  s_{\perp\beta}
\nonumber\\
      && \cdot \frac{2}{16 p_1^+ p_2^-}      \int \frac{d^2 k_\perp}{(2\pi)^2}
       \frac{k_{\perp\alpha} }{k^2_\perp+ i\varepsilon}\frac{1}{ (\vec k_\perp +\vec k_{g\perp} )^2 +(1-x)^2 m^2 }.
\end{eqnarray}
For the physical process, the integration over $k_\perp^2$ is bounded from the above because
the energy-momentum conservation. Since we only work at the leading order of $q^2_\perp$, we
can integrate $k_\perp^2$ from 0 to $\infty$.
For $\delta(k_g^2)$ we can write it with the variables $x$ and $y$ as:
\begin{equation}
   \delta (k_g^2 ) = \delta (2 p_1^+ p_2^- (1-x) (1-y) -q^2_\perp)
    \approx \frac{\delta(1-y)}{2 p_1^+ p_2^- (1-x)} +\cdots, \ \ \ \
       q^- =y p_2^-, \ \ \ \ q^+ = x p_1^+.
\end{equation}
where the terms represented with $\cdots$  will not contribute. Then we have:
\begin{equation}
W_T^{(1)} \vert_{bf} = m \frac{\alpha_s}{\pi^2}  \frac{N_c^2-1}{ N_c}
            \frac{x (1-x)\delta (1-y)  }{ 16 q_\perp^2 (q^2_\perp + (1-x)^2 m^2) }
            \ln \frac {(1-x)^2 m^2 }{q^2_\perp + (1-x)^2 m^2 } .
\end{equation}
\par
There are two cut diagrams for the absorptive part of Fig.3c.
At leading order of $\lambda$ they contributions  cancel
each other. Hence we have the total $W_T^{(1)}$:
\begin{equation}
W_T^{(1)}(x,y,q_\perp^2)  =  m  \frac{\alpha_s^2}{ 8\pi^2} \frac{N_c^2-1}{ N_c}
            \frac{x (1-x) \delta (1-y) }{ q_\perp^2 (q^2_\perp + (1-x)^2 m^2) }
            \ln \frac {(1-x)^2 m^2 }{q^2_\perp + (1-x)^2 m^2 }.
\label{DY}
\end{equation}
With the assignment of power of $\lambda$ in Eq.(19) we find that the leading order of
$W_T^{(1)}$ is of $\lambda^{-4}$. With the above discussion about denominators of propagators,
it seems also possible  that the leading contribution
comes from the region of the loop momentum with $k_\perp << q_\perp$, i.e., the region with the assignment
of power of $\lambda$:
\begin{equation}
  k^\mu = (\lambda^2, \lambda^2, \lambda^2, \lambda^2), \ \ \ \ \ k_g^\mu =( k_g^+, k_g^-, \lambda, \lambda).
\end{equation}
Performing the power counting and evaluating the nominator in Eq.(18) for each diagram
we find the leading contributions from this region to $W_T^{(1)}$ are at order of $\lambda^{-3}$.
Calculating these contributions one finds $W_T^{(1)}=0$. This indicates
that nonzero contribution from this region to $W_T^{(1)}$ starts at order of $\lambda^{-2}$.
One can also perform similar analysis for other regions of the loop momentum.
The conclusion is that the leading term of $W_T^{(1)}$ in $q_\perp$ comes
from the region specified with Eq.(19).
\par
The tree-level TMD parton distribution is:
\begin{equation}
  \bar q(x,k_\perp) = \delta(1-x) \delta^2 (k_\perp) +{\mathcal O}(\alpha_s)
\label{tmd2}  .
\end{equation}
Our results in Eq. (\ref{Sivers},\ref{DY},\ref{tmd2}) shows that the TMD
factorization for $W_T^{(1)}$ in Eq.(\ref{TMDFAC})
is verified at leading order. Although all calculations
performed here are at leading order, they are non-trivial. With the factorization at leading order
the nonzero transverse momentum $q_\perp$ are generated by the nonzero transverse momenta of incoming
partons which are $q$ and $\bar q$. Beyond the leading order, the transverse momentum
$q_\perp$ can also be generated by soft gluon radiation, a soft factor should be included in Eq.(\ref{TMDFAC}).
This soft factor can be identified by extending our calculation to the next-to-leading order as shown
in TMD factorization of $T$-even quantities\cite{CS,CSS,JMY,JMYG}.
\par

\par
\begin{figure}[hbt]
\begin{center}
\includegraphics[width=7cm]{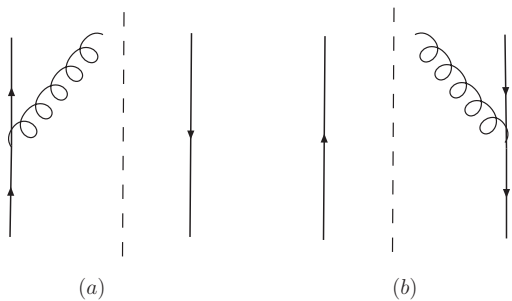}
\end{center}
\caption{The diagrams for the twist-3 matrix element in the $n\cdot G=0$ gauge. }
\label{Feynman-dg3}
\end{figure}
\par

\par\vskip20pt
\noindent
{\bf 5. Twist-3 Matrix Element with a Quark-State}
\par
Now we turn to the twist-3 matrix element. With the single parton state as used for calculating Sivers
function we can also calculate the twist-3 matrix element $T_F(x_1,x_2)$ defined in Eq.(\ref{tw3}).
The calculation can be done easily in the gauge $n\cdot G=0$. In this gauge the gauge links
in Eq.(\ref{tw3}) become a unit matrix in the color-space. The leading order
contribution comes from diagrams given in Fig.5. A straightforward calculation gives:
\begin{eqnarray}
T_F(x_1,x_2)
   = 2\pi \alpha_s C_F m (x_2-x_1)^2 \delta (1-x_2) \int \frac{d^2 k_\perp}{(2\pi)^2}
     \frac{1}{k_\perp^2 + m^2 (1-x_1)^2} + (x_1 \rightleftharpoons x_2).
\end{eqnarray}
The results are U.V. divergent. We can regularize the U.V. divergence
with the dimensional regularization and derive the renormalized $T_F(x_1,x_2,\mu)$:
\begin{equation}
T_F(x_1,x_2,\mu)
   =  \frac{\alpha_s}{2} C_F m (x_2-x_1)^2 \delta (1-x_2) \ln \frac{\mu^2}{(1-x_1)^2 m^2} + (x_1 \rightleftharpoons x_2).
\end{equation}
We note that $T_F(x_1,x_2)$ is zero with $x_1=x_2$ at leading order and the renormalization scale
$\mu$ acts effectively as a cutoff of the transverse momentum.
\par
In the collinear factorization of SSA in Eq.(\ref{CF}), the leading order of the perturbative
function $H_c$ is at $\alpha_s$ and $H_c$ does not contain explicit $\mu$-dependence. With
the calculated $T_F$ and
leading order results of the standard parton distribution one can find
that the collinear factorization of SSA with the twist-3 matrix element fails to reproduce
the partonic SSA in Eq.(\ref{DY}) completely.
In Eq.(\ref{CF}) the contributions can  be divided into the contributions from soft-poles of propagators
and hard-poles of propagators. The contribution of soft poles is proportional to
$T_F(x,x)$. Because $T_F(x,x)$ with our partonic state is zero at leading order of $\alpha_s$,
we can not verify the soft-pole contribution with our results.
With our results one can actually derive a factorized formula with $x<1$ and $\mu =q_\perp$ as:
\begin{eqnarray}
W_T^{(1)}(x,y,q_\perp) &=& -
  \frac {\alpha_s}{ 2 \pi^2 (q^2_\perp)^2} \int \frac{dy_1}{y_1} \frac{ d y_2}{ y_2} \bar q(y_2) \delta (1- \xi_2)
     \frac{ 1 }{(1-\xi_1)_+} x T_F(x,y_1,q_\perp),
\end{eqnarray}
with $\xi_1 = x/y_1$, $\xi_2 =y/y_2$ and $\bar q(y_2)$ as the standard antiquark distribution.
There is a certain ambiguity to determine the hard kernel in the above. Because $T_F(x,x,q_\perp)=0$
one can replace the above $+$-distribution with $1/(1-\xi_1)$.
The above factorized contribution
only corresponds to a part of the hard-pole contribution in the approach of the collinear factorization.
The above factorization has a clear meaning in physics. In Fig.3b the gluon emitted by the quark
is collinear to the quark in the limit $q^2_\perp << Q^2$. This collinear system
is factorized into the twist-3 matrix element $T_F$.
Our study also indicates that the relation between
$q_\perp(x,k_\perp)$ and $T_F(x_1,x_2)$ for $k_\perp^2 >> \Lambda^2_{QCD}$, derived formally in \cite{JQVY},
is not satisfied with our partonic results.
It will be interesting for a further
study to show how one can find all contributions in the approach of the collinear factorization
with some partonic states.
\par
There is another relation between $q_\perp$ and $T_F$ derived formally
in \cite{Boer03,MW1}:
\begin{equation}
   T_F(x,x) = \int d^2 k_\perp k_\perp^2 q_\perp (x, k_\perp).
\end{equation}
With our result of Sivers function in Eq.(\ref{Sivers}) one can realize that the integral of $k_\perp$
is U.V. divergent. Therefore the integral without any subtraction is not well-defined. A possible way
to modify the integral meaningful in order to relate Sivers function with $T_F$ is to define:
\begin{eqnarray}
  Q_\perp (x,b) &=& \int d^2 k_\perp k_\perp^2 q_\perp (x, k_\perp) \exp(-i\vec b \cdot \vec k_\perp )
\nonumber\\
  &=& -\frac{ m \alpha_s^2}{16 \pi}  (N_c^2-1)  x(1-x) \ln^2 \left [ \frac{(1-x)^2 m^2 b^2 e^{2\gamma}}{4} \right ]
    + {\mathcal O}(b^2),
\end{eqnarray}
where we have used our result in Eq.(\ref{Sivers}) to produce the expression in the second line.
One may expect that $T_F(x,x)$ is somehow related to Sivers function in the impact space. We examine this in
the below.
\par
\begin{figure}[hbt]
\begin{center}
\includegraphics[width=10cm]{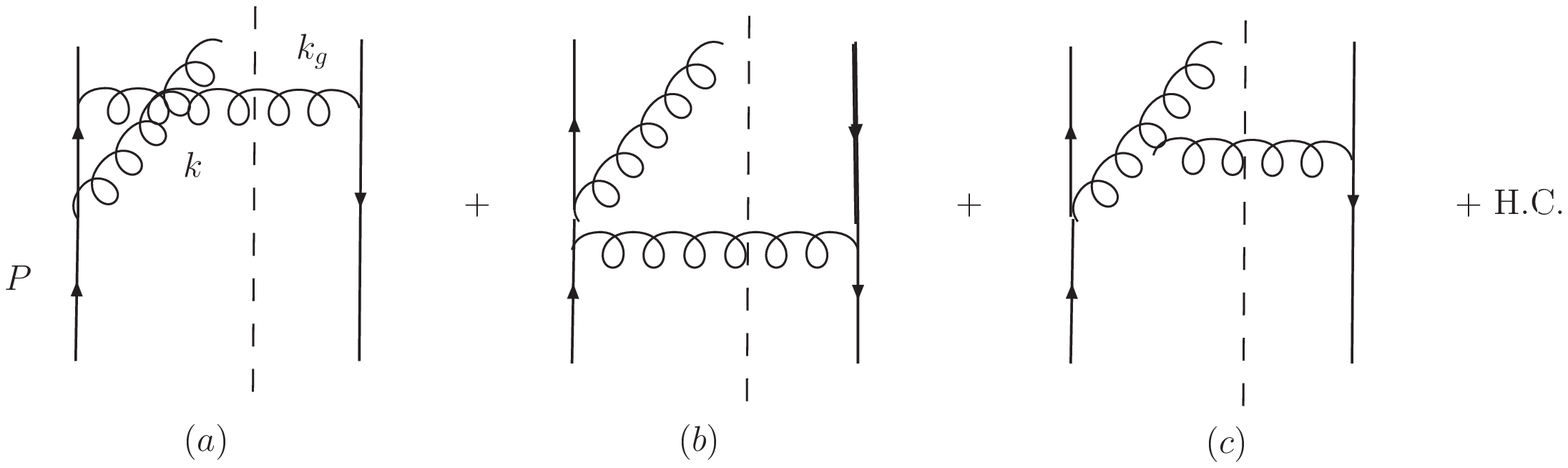}
\end{center}
\caption{Diagrams for the contributions to $T_F(x,x)$. }
\label{Feynman-dg6}
\end{figure}
\par
\par
We have seen at leading order $T_F(x,x)=0$. This function can be nonzero
at higher orders where a gluon can be in the intermediate state, i.e., a gluon crosses
the cut in Fig.5. These diagrams are given in Fig.6.
For $x_1=x_2$ one can easily find that
only the contribution from Fig.6c is not zero. The calculation is similar as before. We find
\begin{eqnarray}
T_F(x_1,x_2)  s_{\perp}^{\nu}\vert_{6c} &=&
    -i g_s^4 (x_2 -x_1) p^+ \frac{  f^{abc} {\rm Tr} T^a T^b T^c }{N_c} \int \frac{ d^4 k}{(2\pi)^4}
   \frac{ d^4 k_g}{(2\pi)^4} (2\pi) \delta ( k_g^2)
\nonumber\\
  &&\cdot  \pi \delta (x_1 p^+ -(p^+ -k_g^+ -k^+))
       \delta (k^+ -(x_2-x_1)p^+) \epsilon_\perp^{\mu\nu}
\nonumber\\
  && \cdot \frac{1} {k^2 + i\varepsilon} \frac{1} {(k +k_g)^2 + i\varepsilon}
     \frac{1} {(p-k_g)-m^2 - i\varepsilon}\frac{1} {(p-k_g-k)^2 -m^2 + i\varepsilon}
\nonumber\\
   && \cdot 8 i m k_{\perp\mu} (-1) \epsilon_\perp^{\alpha \beta} k_{\perp\alpha} s_{\perp\beta}
          \frac{ k_g^+ }{k^+}\left (p^+ -k_g^+ +\frac {\epsilon}{2} k_g^+  \right ) \left [ 1 +{\mathcal O}(k^+) \right ].
\end{eqnarray}
Because we will encounter U.V. divergences we have used the naive $\gamma_5$-prescription in $d=4-\epsilon$
dimension. The term with $\epsilon$ in the last line comes from the trace of $\gamma$-matrices.
Performing integrations and subtracting U.V. divergence we have for $T_F(x_1,x_2)$ with $x_1=x_2 =x$:
\begin{eqnarray}
T_F(x,x) &=& - \frac{m \alpha_s^2}{4\pi}(N_c^2-1)(1-x)  x
 \left ( \ln^2 \frac{\mu^2}{(1-x)^2 m^2} + \ln \frac{\mu^2}{(1-x)^2 m^2}
        + \frac{\pi^2}{12} + \frac{1}{2} \right )
\nonumber\\
    &&     - \frac{m \alpha_s^2}{4\pi}(N_c^2-1) (1-x)^2 \left ( \frac{1}{2} + \ln \frac{\mu^2}{(1-x)^2 m^2}\right ) .
\end{eqnarray}
From the above results it is clear that one can in general not write down a factorized relation like
\begin{equation}
T_F(x,x, \mu) =  C(x,\mu, b) \int d^2 k_\perp k_\perp^2 q_\perp (x, k_\perp) \exp(-i\vec b \cdot \vec k_\perp )
                 + {\mathcal O} (b)
\end{equation}
with the coefficient $C(x,\mu, b)$ as a perturbatively calculable coefficient.
\par\vskip20pt
\noindent
{\bf 6. Summary}
\par
SSA is a $T$-odd effect and it requires helicity-flip interactions.
Two factorization approaches to study SSA has been suggested. One approach is by using TMD factorization
which takes transverse momenta of partons into account. In this approach the $T$-odd effect and
the effect of helicity-flip interactions are parameterized by Sivers function, which can consistently
be defined with QCD operators. Another approach is to use standard collinear factorization. In this approach
the effect of helicity-flip interactions is parameterized by the twist-3 matrix element, while the $T$-odd
effect arises from hard scattering of partons. Sofar, all factorized formulas of SSA in the two approaches
have been derived in a rather formal way  in the following sense: One has used the diagram expansion
with hadrons and has divided a given diagram with hadrons into various parts.
Among these various parts, one consists only of partons. In other parts hadrons are involved.
These parts represent nonperturbative effects related to the hadrons and they are parameterized
by various parton density matrices of hadrons. Since a proven factorization is in general a property
of QCD, the above two factorizations should hold by replacing hadrons with partons, if the two
can be proven. The factorization approaches have been not examined with partonic states.
\par
The study of our work presented here is to examine the two factorizations of SSA in Drell-Yan processes
with partonic states in first time. We replace the initial hadrons with a quark and an antiquark, where
the quark is transversely polarized. With the quark state we can calculate Sivers function and
twist-3 matrix element. SSA can be calculated with the quark-antiquark state. The finite quark mass
is introduced to flip helicities. With our partonic results we can examine the two factorization
approaches at leading but nontrivial order of $\alpha_s$. It turns out that our partonic result
of SSA can be factorized within the approach of TMD factorization, but the results of
the proposed factorization formula of the collinear factorization can not be verified completely
with our partonic results.
This is our main result. Using our results we have also examined
two formally derived relations between Sivers function and the twist-3 matrix element.
Our partonic results do not satisfy these two relations.
It will be interesting to see how one can derive the formally derived results
in the collinear factorization with some partonic results.
\par
In our work we have taken a quark-antiquark state to replace the initial hadrons in Drell-Yan
processes. We can extend our study to the case of a quark-gluon state to examine the two approaches.
It is worth to point out that corrections of the perturbative coefficient at higher orders of $\alpha_s$
can be studied after the verification of a factorization with partonic states at leading order. For SSA
the verification at leading order is nontrivial as shown in this work. With our progress
to understand SSA in Drell-Yan processes,
we are able to study SSA in semi-inclusive DIS and other possible processes.
Such a study is currently under the way.

\par\vskip20pt
\par\noindent
{\bf\large Acknowledgments}
\par
We thank Prof. X.D.Ji, J.W. Qiu and F. Yuan for helpful discussions.
This work is supported by National Nature Science Foundation of P.R. China((No. 10721063,10575126).
\par

\end{document}